\documentclass[journal]{IEEEtran}

\IEEEoverridecommandlockouts
\usepackage{booktabs}

\usepackage{longtable}
\usepackage[numbers]{natbib}
\usepackage{booktabs}
\usepackage{amsmath,amssymb,amsfonts}
\usepackage{algorithmic}
\usepackage{graphicx}
\usepackage{multicol}
\usepackage{booktabs}
\usepackage{multirow}
\usepackage{textcomp}
\usepackage{smartdiagram}
\usesmartdiagramlibrary{additions}
\usepackage{booktabs}
\usepackage{graphicx}
\usepackage{lscape}
\usepackage{rotating}
\usepackage{longtable}
\usepackage{url}
\usetikzlibrary{shapes.geometric, arrows}
\usepackage{hyperref}

\usepackage{titlesec}
\titleclass{\subsubsubsection}{straight}[	hesubsubsection]
\newcounter{subsubsubsection}[subsubsection]
\renewcommand\thesubsubsubsection{\thesubsubsection.\arabic{subsubsubsection}}
\titleformat{\subsubsubsection}[runin]{\normalfont\normalsize\bfseries}{\thesubsubsubsection}{1em}{}
\titlespacing*{\subsubsubsection}{0pt}{0.5em}{0.5em}

\usepackage{tikz}
\usetikzlibrary{shapes.geometric, arrows}

\tikzstyle{startstop} = [rectangle, rounded corners, text width=2cm, minimum height=1cm, text centered, draw=black, fill=white]
\tikzstyle{process} = [rectangle, text width=2.5cm, minimum height=1cm, text centered, draw=black, fill=white]
\tikzstyle{decision} = [diamond, text width=2cm, minimum height=0.8cm, text centered, draw=black, fill=white, aspect = 2]
\tikzstyle{arrow} = [thick,->,>=stealth]

\usepackage{xcolor}
\def\BibTeX{{\rm B\kern-.05em{\sc i\kern-.025em b}\kern-.08em
    T\kern-.1667em\lower.7ex\hbox{E}\kern-.125emX}}
\begin{document}

\title{Prompt Engineering for Large Language Model-assisted Inductive Thematic Analysis}

\author{Muhammad Talal Khalid and
        Ann-Perry Witmer
\thanks{M. T. Khalid is in the Department
of Electrical and Computer Engineering, University of Illinois Urbana Champaign, Urbana,
IL, 61801 USA e-mail: (mkhalid4@illinois.edu).}
\thanks{A-P. Witmer is associated with the Carle Illinois College of Medicine, University of Illinois Urbana Champaign, Urbana,
IL, 61801 USA e-mail: (awitmer@illinois.edu).}
}

\maketitle

\maketitle

\begin{abstract}
The potential of large language models (LLMs) to mitigate the time- and cost- related challenges associated with inductive thematic analysis (ITA) has been extensively explored in the literature. However, the use of LLMs to support ITA has often been opportunistic, relying on ad hoc prompt engineering (PE) approaches, thereby undermining the reliability, transparency, and replicability of the analysis. The goal of this study is to develop a structured approach to PE in LLM-assisted ITA. To this end, a comprehensive review of the existing literature is conducted to examine how ITA researchers integrate LLMs into their workflows and, in particular, how PE is utilized to support the analytical process. Built on the insights generated from this review, four key steps for effective PE in LLM-assisted ITA are identified and extensively outlined. Furthermore, the study explores state-of-the-art PE techniques that can enhance the execution of these steps, providing ITA researchers with practical strategies to improve their analyses. In conclusion, the main contributions of this paper include: (i) it maps the existing research on LLM-assisted ITA to enable a better understanding of the rapidly developing field, (ii) it outlines a structured four-step PE process to enhance methodological rigor, (iii) it discusses the application of advanced PE techniques to support the execution of these steps, and (iv) it highlights key directions for future research.

\end{abstract}


\section{Introduction}

\label{Sec 1}

The advent of generative artificial intelligence (AI), particularly large language models (LLMs), has profoundly transformed various research domains. From healthcare to engineering design, and from law to politics, it has impacted the day-to-day operations of society in ways that were unheard of before \cite{kaddour2023challenges}. The more recent advancements of these models have made them cost-effective and even more accessible and available \cite{gibney2025china}. Following this curve, inductive thematic analysis (ITA) researchers have also begun to explore the application of LLMs to support their analyses. From cautious employment of LLMs as a research assistant to the proposal and development of stand-alone computer applications that can (potentially) conduct ITA with a few clicks \cite{zhang2023qualigpt}, the integration of these models into the ITA research is becoming pervasive.

LLMs are specialized generative AI models that process natural language and are trained on `large' amounts of textual data. They use advanced machine learning techniques (e.g., deep learning) to learn from data, understand context, identify complex patterns, and respond to user queries accordingly \cite{kaddour2023challenges}. These capabilities enable LLMs to engage in meaningful conversations and assist in tasks that previously required exclusively human intervention \cite{park2023thinking}. There are, however, limitations that constrain the effective use of LLMs. The proprietary nature of advanced LLMs has met criticism on ethical grounds, focusing particularly on concerns related to data privacy and security \cite{schroeder2024large}. In addition, the completeness, recency, and breadth of the data used to train these models also casts doubt on the truthfulness of the model responses and can lead to biases towards specific social groups and viewpoints \cite{khan2024automating}. Moreover, the LLMs' notorious power of expression and suggestibility render them prone to hallucinations, i.e., to generate fabricated data and arguments that are misrepresented as the truth \cite{schroeder2024large}.

Even with all these limitations, LLMs offer complementary analytic power that qualitative researchers can leverage to critically evaluate and engage with their data. In this domain, ITA is a highly flexible and widely used systematic data analysis method to identify, analyze, and interpret patterns of meaning within qualitative data \cite{zhang2023redefining}. It is a flexible tool often employed by a wide range of qualitative researchers to understand experiences, thoughts, or behaviors represented in a dataset \cite{kiger2020thematic}. Unlike deductive approaches of qualitative inquiry in which the data are labeled according to pre-specified themes, ITA's bottom-up focus employs the data itself to identify emerging and interesting common ways in which topics in the data are represented. 

Often ITA researchers follow Braun and Clarke’s six-step process to conduct their analyses \cite{braun2024thematic}. According to this process, researchers first familiarize themselves with the data through repeated reading. They then generate initial codes and identify potential themes in the data. These themes are reviewed, refined, defined, named, renamed, and are finally integrated into a structured report to present the findings. As a result, performance of ITA is both costly and time-consuming, particularly when it is used to analyze large datasets. In addition, establishment of credibility requires the involvement of at least two trained researchers with relevant expertise, who must engage in multiple rounds of discussions. This labor-intensive process further increases the overall cost \cite{zhao2024new}.

In this regard, the ability of LLMs to synthesize large volumes of data efficiently, enabled by their high processing speed, makes them well-suited to address some limitations of ITA. Although the integration of LLMs in ITA is still in a nascent state, there is a growing interest among researchers in their use \cite{de2024performing}. Multiple studies have compared LLM-aided ITA to the traditional counterparts to assess reliability and establish confidence in their use (see Section \ref{Sec 3}). The similarity between the two analyses reported in these studies underscores the promising contributions that LLMs can offer to the ITA field. As some researchers observe, `` ... it seems inevitable that ITA researchers will need to engage with LLMs to enhance their work" \cite{de2024performing}.

Some earlier concerns regarding the use of LLMs to support ITA have already been partially or completely addressed in the recent, more advanced LLMs. These include concerns related to the limited length of textual data that can be inputted to the model, the memory-less operation of the model, the need to upload qualitative data to cloud-computing resources that threatens data privacy and security, and the costly operations that accompany the use of these models \cite{de2024performing, mannstadt2024novel, deiner2024large}. But the more recent LLMs can take 1 million tokens\footnote{Tokens are the fundamental units of language that represent words, phrases, or parts of words} and therefore remember longer conversations \cite{pichai2024Ournext}. These models can be used on local devices and are becoming less costly to operate \cite{gibney2025china}. Yet other concerns such as those related to the training data limitations, the ethics of using LLMs, and the model hallucinations as previously mentioned require careful consideration, and deliberations elsewhere should be fully supported.

In addition, another issue that has been repeatedly discussed by multiple ITA researchers is the crucial importance of prompts that are used to generate LLM responses. Since LLMs function by probabilistically predicting the next word in a sequence based on the input they receive, a slight variation in the language used in prompts can significantly influence the quality of the model's responses. A recent study reported a significant shift in the attitudes of seventeen junior qualitative researchers from skepticism to confidence in using LLMs as they became more knowledgeable about effective mechanisms to interact with these models \cite{zhang2023redefining}. Thus, a carefully designed and systematically refined prompt can lead to more accurate, relevant, and coherent responses, effectively guiding the model to meet the intended objectives. 

The process of strategically crafting and iteratively refining task-specific prompts to optimize an LLM's responses is referred to as Prompt Engineering (PE) \cite{schulhoff2024prompt}. While ITA researchers have employed various approaches to PE, there remains a critical need for a systematic examination and synthesis of these methods.This paper aims to address this gap. As such, the key contributions of this work include:
\begin{itemize}
    \item Through a systematic review of the literature, this study provides an overview of the use of LLMs for ITA and the particular PE techniques used by ITA researchers.
    \item Built on the insights generated from the literature review, this article proposes a four-step PE process to enhance methodological rigor and effectiveness of LLM-assisted ITA.
    \item The study then discusses the application of advanced PE techniques in the ITA context to assist researchers in executing the four PE steps.
\end{itemize}


The remainder of this paper is structured as follows. Section \ref{Sec 2} concisely introduces PE. Section \ref{Sec 3} presents a systematic review of the literature and explores how PE is approached by ITA researchers in their studies. The four-step PE process is outlined in Section \ref{Sec 4}. A brief description of state-of-the-art PE techniques to support the four PE steps is provided in Section \ref{Sec 5}. The paper concludes with directions for future work in Section \ref{Sec 6}.

\section{An Introduction to Prompt Engineering}

\label{Sec 2}


In the LLM terminology, a prompt is ``a collection of instructions that are provided to a model to guide its responses to a particular task". These instructions can take the form of text, images, voice recordings, other media, or any combination thereof. Numerous researchers have devoted significant efforts to identify and elucidate the key components of a prompt. According to \cite{marvin2023prompt} and \cite{dair2025}, a prompt consists of researcher-defined instructions and may include input data, task context, and output indicators. \cite{IBM2025} adds task-specific examples to the list of prompt components to emphasize the importance of demonstrative examples that illustrate the reasoning the researcher wants the LLM to follow. Moreover, \cite{Google2025} highlights the explicit definition of the role the researcher expects the model to assume as a critical prompt component. On the other hand, \cite{schulhoff2024prompt} identifies six key elements of a prompt: the directive to the model in the form of an instruction or a question, style instructions to guide the model on how the researcher desires model response to look and feel, the structure and formatting of the response, the role the model should take when responding to a directive, examples demonstrating how the model should accomplish the task, and additional background or external information to provide clarity. Prompt elements reported in these studies can be synthesized into the following four:  


\begin{table}[]
\centering
\resizebox{\columnwidth}{!}{%
\begin{tabular}{@{}p{3cm}p{1cm}p{1cm}p{1cm}p{1cm}@{}}
\toprule
  \textbf{Prompt} &
  \textbf{Task directive} &
  \textbf{Exemplar} &
 \textbf{ Framing} &
  \textbf{Response Instructions} \\

  \midrule

  Identify most relevant themes in the following text: \{text to be analyzed\}& Theme identification &
  - &
   &
  - \\ 
  
  \midrule

  Generate a very broad range of initial codes in the following text. Provide a name for each code in no more than 4 words with 40 words of meaningful and dense description. \{text to be analyzed\} &
  Theme naming &
  - &
  - &
  Response word limit \\
  
  \midrule

  The attached file provides two examples of how to generate initial codes. Refer to these examples and generate codes in the following text: \{text to be analyzed\}  &
  Code generation &
  Initial coding examples &
  - &
  - \\
  
  \midrule
  You are a content analysis bot. You analyze content from transcripts. Given this information, please identify the main themes in the text below: \{text to be analyzed\}&
  Theme identification &
  - & Role specification, data description
   &
  \\ 
  
  \bottomrule
  \\

\end{tabular}%
}
\\
\caption{Prompt Examples.}
\label{Prompt examples}
\end{table}

\begin{itemize}
    \item \textbf{Task directive:} The implicit or explicit instructions or questions to which the researcher wants the model to respond. 
    \item \textbf{Exemplars:} Illustrative patterns or logic that the researcher wants the model to follow when it responds to the task directive.
    \item \textbf{Framing}: Background information provided to the model to help it better understand the task context. Prompt framing may also include any input data and its explanation that would be required by model to complete the task.
    \item \textbf{Response instructions:} Guidelines specifying the desired constituents, structure, formatting, and style of the model's response.
\end{itemize}

Table \ref{Prompt examples} highlights the various prompt elements in prompt examples extracted from articles reviewed in \ref{Sec 3}. Note that, with the exception of the task directive, all other prompt elements are optional.

Prompt engineering (PE) is the process of strategically crafting and iteratively refining task-specific prompts to optimize an LLM's responses \cite{schulhoff2024prompt}. Thus, the two important aspects of PE include crafting task-specific prompts and refining the outputs generated by the model until the desired result is achieved. While various PE best practices are explained in the literature, a key shortcoming is the lack of systematic reporting on how ITA researchers have approached PE in their LLM-aided analyses. This is important to improve reliability, replicability, and transparency of the conducted research \cite{shah2024prompt}. To this end, the following section presents a review of the use of PE in the LLM-aided ITA literature.

\section{Systematic Literature Review}

\label{Sec 3}

 This section presents a systematic literature review of the field and comprehensively analyzes the collected data. The review was carried out in two main steps, i.e., planning and conducting, following the guidelines presented in \cite{carrera2022conduct}.

\subsection{Planning}
In the planning step, a clear research direction was identified and research questions were formulated. Digital databases used for article search were selected, article selection criteria were defined, and a data extraction form was created. The two research questions that this systematic review aims to address include:

\begin{itemize}
    \item \textbf{RQ1:} How have LLMs been used to assist ITA? \\
   \textbf{ Rationale:} Before examining PE approaches in the ITA literature, it is essential to understand the research context and objectives of LLM-aided ITA. This question helps in understanding the qualitative data analyzed, the analysis field, and the criteria used to evaluate LLMs' ITA performance.
    
    \item \textbf{RQ2:} How have researchers approached PE when using LLMs to support ITA? \\
    \textbf{Rationale:} This question seeks to identify the predominant approaches to PE used by ITA researchers. Understanding this is essential to highlight common practices in the ITA community and to highlight the opportunities for ITA researchers to improve their confidence in and the quality of LLM-aided ITA.
    
\end{itemize}

Once the research questions were defined, a comprehensive article search strategy was devised. Due to the multidisciplinary nature of investigation (application of technical models to mostly non-technical research), an interdisciplinary database (Scopus\footnote{\url{https://www.scopus.com/}}), an engineering database (IEEE Digital Library\footnote{\url{https://ieeexplore.ieee.org/Xplore/home.jsp}}), and a computing and information technology database (ACM Digital Library\footnote{\url{https://dl.acm.org/}}) were selected. To account for the recent nature of the topic under investigation and the possibility of studies still awaiting peer review, the Arxiv database\footnote{\url{https://arxiv.org/}} was additionally consulted. 

Subsequently, selection criteria were established to support the proper identification of relevant studies for this literature review. These selection criteria included assessment of primary bibliographic data as well as examination of articles to decide their relevance to the research questions. The four selection criteria (SC) used for this study are reported in Table \ref{tab:selectionCriteria}. These criteria were applied using boolean logic with the possibility of yes or no responses only.

\begin{table}[]
\centering
\resizebox{\columnwidth}{!}{%
\begin{tabular}{@{}p{1cm}p{5.5cm}p{1cm}@{}}
\toprule
        \textbf{Inclusion Criteria} & \textbf{Description} & \textbf{Response Options}\\
\midrule
        SC1& The study is a journal article, conference article or a poster & Yes/No\\
        SC2& Either the title, keywords, or abstract of the article contains the search string keywords & Yes/No\\
        SC3& The study analyzes the application of LLMs to support ITA & Yes/No\\
        SC4& The article includes the prompts used to conduct ITA & Yes/No \\ \bottomrule
\end{tabular}%
}
\vspace{0.5cm}
\caption{Selection Criteria.}
\label{tab:selectionCriteria}
\end{table}

A data extraction form, which allows free-form responses, was created to synthesize the information necessary to answer the two research questions. Table \ref{tab:dataextractionform} presents the data extraction form and maps each question on the form to the corresponding research question. 

\begin{table}[]
\centering
\resizebox{\columnwidth}{!}{%
\begin{tabular}{@{}p{1cm}p{1cm}p{6cm}@{}}
\toprule
\textbf{RQ} & \textbf{DEQ} & \textbf{Description}                                                 \\ \midrule
\multirow{4}{*}{RQ1} & DEQ1 & What is the context in which the study is conducted and what data are used?                \\
                    & DEQ2 & What are the primary objectives and limitations of the study ?                              \\
                     & DEQ3 & What LLM was used to conduct the analysis?                            \\
                     & DEQ4 & How does the study evaluate the performance of LLM in aiding ITA?                                   \\
                               \midrule
\multirow{4}{*}{RQ2}            & DEQ5 & How does the study define the prompts used to conduct LLM-assisted ITA? \\
                     & DEQ6 & Does the study explain how model responses were refined?\\
                     \bottomrule
\end{tabular}%
}
\vspace{0.5cm}
\caption{Data Extraction Questions.}
\label{tab:dataextractionform}
\end{table}

\subsection{Conducting}

After defining the ground rules, the systematic review process entered the conducting step. In this phase, the digital library search string was iteratively formulated and used to extract articles using SC. The data extraction form was filled with relevant information which was then analyzed to draw key insights. 

A generic search string was developed using keywords ``large language models" and ``thematic analysis". These keywords were connected using the logical operator ``AND", while their synonyms were linked with the ``OR" operator. These search string terms were selected keeping in view the research questions for this systematic review. The search string was tested in various configurations on the Arxiv database, and a consensus agreement between the authors resulted in the definition of the final version, i.e.,


\begin{quote}
    (``LLM" OR ``LLMs" OR ``Large Language Model" OR ``Large Language Models") AND (``Thematic Analysis" OR ``Thematic Analyses")
\end{quote}

\tikzstyle{startstop} = [rectangle, rounded corners, text width=2cm, minimum height=1cm, text centered, draw=black, fill=white]
\tikzstyle{process} = [rectangle, text width=3cm, minimum height=1cm,  draw=black, fill=white, minimum width = 2cm]
\tikzstyle{decision} = [diamond, text width=2cm, minimum height=0.8cm, text centered, draw=black, fill=white, aspect = 2]
\tikzstyle{arrow} = [thick,->,>=stealth]

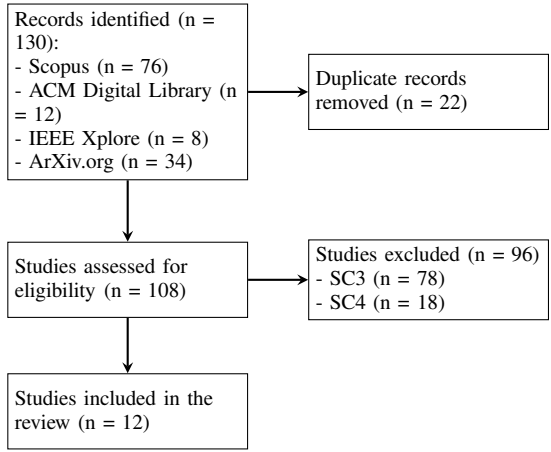
\begin{figure}
    \centering
    \footnotesize
    \begin{tikzpicture}
    
    \node (step1) [process] {Records identified (n = 130): \\
    - Scopus (n = 76)\\
    - ACM Digital Library (n = 12) \\
    - IEEE Xplore (n = 8) \\
    - ArXiv.org (n = 34)};
    
    \node (int1) [process, right of=step1, xshift=3cm] {Duplicate records removed (n = 22)};

    \node (step2) [process, below of=step1, yshift=-1.5cm] {Studies assessed for eligibility (n = 108)};
    
    \node (int2) [process, right of=step2, xshift=3cm] {Studies excluded (n = 96)\\
    - SC3 (n = 78)\\
    - SC4 (n = 18)};

    \node (step3) [process, below of=step2, yshift=-0.75cm] {Studies included in the review (n = 12)};

    
    \draw [arrow] (step1) -- (int1);
    \draw [arrow] (step1) -- (step2);
    \draw [arrow] (step2) -- (int2);
    \draw [arrow] (step2) -- (step3);

    \end{tikzpicture}
    \caption{Study selection process.}
    \label{fig:Studyselection}
\end{figure}

Figure \ref{fig:Studyselection} illustrates the steps followed in the study selection process. The execution of the final search string and the application of SC1 and SC2 resulted in the retrieval of a total of 130 studies from the four databases (Scopus: 76, ACM Digital Library: 12, IEEE Digital Library: 8, Arxiv: 34). The searches were carried out in the months of January and February 2025. A further refinement process was employed to remove duplicates resulting in a 108 studies, while the application of SC3 and SC4 reduced the number of final studies to be analyzed to 12 articles which are indexed in Table \ref{tab:selectedarticles}.

\begin{table}[]
\resizebox{\columnwidth}{!}{%
\begin{tabular}{@{}p{1cm}p{6.5cm}p{1cm}@{}}
\toprule
\textbf{ID} &
  \textbf{Study Title} &
  \textbf{Ref.} \\ \midrule
A1 &
  Large Language Models can Enable Inductive Thematic Analysis of a Social Media Corpus in a Single Prompt: Human Validation Study &
  \cite{deiner2024large} \\ \midrule
A2 &
  Large Language Model Versus Human-Generated Thematic Analysis in Otolaryngology Qualitative Research &
  \cite{morse2025large} \\ \midrule
A3 &
  Comparing the efficacy and efficiency of human and generative AI: Qualitative thematic analyses &
  \cite{prescott2024comparing} \\ \midrule
A4 &
  LLM-in-the-loop: Leveraging Large Language Model for Thematic Analysis &
  \cite{dai2023llm} \\ \midrule
A5 &
  Inductive Thematic Analysis of Healthcare Qualitative Interviews Using Open-source Large Language Models: How Does it Compare to Traditional Methods? &
  \cite{mathis2024inductive} \\ \midrule
A6 &
  Thematic Analysis with Open-Source Generative AI and Machine Learning: A New Method for Inductive Qualitative Codebook Development &
  \cite{katz2024thematic} \\ \midrule
A7 &
  Performing an Inductive Thematic Analysis of Semi-structured Interviews with a Large Language Model: An Exploration and Provocation on the Limits of the Approach &
  \cite{de2024performing} \\ \midrule
A8 &
  Using Generative Text Models to Create Qualitative Codebooks for Student Evaluations of Teaching &
  \cite{katz2024using} \\ \midrule
A9 &
  A Comparison of the Results from Artificial Intelligence-based and Human-based Transport-related Thematic Analysis &
  \cite{carvalho2024comparison} \\ \midrule
A10 &
  A Novel Approach for Mixed-Methods Research Using Large Language Models: A Report Using Patients’ Perspectives on Barriers to Arthroplasty &
  \cite{mannstadt2024novel} \\ \midrule
A11 &
  Exploring the Synergy of Human and AI-driven Approaches in Thematic Analysis for Qualitative Educational Research &
  \cite{sabbaghan2024exploring} \\ \midrule
A12 &
  Further Explorations on the Use of Large Language Models for Thematic Analysis. Open-Ended Prompts, Better Terminologies and Thematic Maps &
  \cite{de2024further} \\ \bottomrule
  \vspace{0.25cm}
\end{tabular}%
}
\caption{List of Selected Articles}
\label{tab:selectedarticles}
\end{table}

\subsection{Results}
11 of the 12 selected studies were published in 2024 while the remaining one article was published in 2023. This data highlights the topic's recent emergence and the increasing interest within the ITA community. The results from the data extraction form are presented in the following subsections.

\subsubsection{DEQ1: What is the context in which the study is conducted and what data are used?}
ITA researchers utilized LLMs across various research fields, including healthcare (A1–A3, A5, A10), education (A6-A7, A11-A12), transportation (A9), and organizational (A8) and customer satisfaction research (A4).

In terms of data sources, six selected studies examined semi-structured interviews (A2, A5–A6, A10–A12), three focused on open-ended survey responses (A4, A7–A9), A1 analyzed social media posts, and A3 studied text messages.

\subsubsection{DEQ2: What are the primary objectives and limitations of the study ?}

The selected studies had one of following three primary objectives. Some studies assessed the feasibility of using LLMs to support ITA by comparing the similarity between themes generated by the model and those produced by human researchers (A1-A3, A5, A8-A10). Others proposed experimental frameworks to facilitate ITA with the support of LLMs (A4, A6-A8), while the primary goal of A12 was to develop and test different prompts for initial coding and theme generation for ITA.

The findings in all 12 studies indicate a high level of performance by LLMs in supporting ITA. The themes generated by LLMs generally aligned closely with those produced by human researchers. Notably, A4 observed that the similarity between themes generated by human researchers and those generated by LLMs was even greater than the similarity between themes produced by two independent human researchers.

Despite their effectiveness, the most significant limitation of using LLMs for ITA described in the selected studies is the reproducibility of results (A1-A2, A4, A6, A12). It is highlighted that even minor variations in prompt wording can lead to substantial differences in the quality and quantity of generated codes and themes. Additionally, concerns were also raised regarding the model’s performance in deeper interpretive tasks, such as understanding contextual nuances and recognizing the significance of key insights in less frequent data (A3, A5, A7-A12). According to these studies, because LLMs operate independently of any specific research domain, their identification of initial codes and themes lacked the depth of human interpretation.

\subsubsection{DEQ3: What LLM was used to conduct the analysis?}
Most studies utilized a single LLM, though some tested multiple models to compare the ITA results. The most frequently used LLMs was GPT-3.5 (A3-A4, A6, A9, A12) followed by GPT-4 (A1–A2, A10-A11). Other studies also incorporated Claude (A1), LLAMA2 (A5), Mistral (A8), and Gemini (A3, A9) to support their analyses. No specific data on LLM usage was reported in (A7).

\subsubsection{DEQ4: How does the study evaluate the performance of LLM in aiding ITA?}
All 12 studies compared the ITA conducted with LLM support to the traditional analysis methods. Eleven of the 12 studies (A1-A7, A9-A12) assessed the similarity between themes generated by trained human researchers and those produced by LLMs. On the other hand, A8 examined whether LLM-generated themes aligned with those used to create the hypothetical dataset on which the LLM performed ITA. 

Similarity was determined through manual inspection by researchers or by applying semantic similarity metrics such as cosine similarity (A4). Some studies also incorporated the use of the Jaccard similarity coefficient (A5) and the inter-rater reliability to validate the LLM generated ITA outcomes (A11). Additionally, two studies (A1-A2) also engaged subject matter experts to evaluate the reasonableness of LLM-generated ITA outcomes.


\subsubsection{DEQ5: How does the study define the prompts used to conduct LLM-assisted ITA?}

The prompt definition in the studies analyzed in this work differed in at least two key ways. First, the studies took different approaches to complete LLM-assisted ITA. Second, the components of the prompts used in these studies varied considerably.

As such, five studies (A1-A3, A9, A11) adopted a global approach by instructing the model to complete all ITA steps using a single prompt. The remaining studies (A4–A8, A10, A12) divided the ITA process into multiple sub-tasks, following Braun and Clarke's method. Out of these, except for two studies (A6, A8), all others completed only three out of the six steps in the method using LLMs. The steps conducted with LLM support included identification of initial codes, themes generation, and naming and description of these themes. To assist the model in reaching desired responses, in A6 and A8, the authors provided a step-by-step process for each sub-task that the model should follow, thereby further reducing the complexity. 

Interestingly, four studies (A1, A5, A9, A11) additionally experimented with multiple prompts to generate model responses when completing the ITA task (or multiple sub-tasks of the Braun and Clarke's method). These responses were then aggregated to reach the final outputs. This experimentation involved running the same prompt multiple times with identical content while setting a non-zero prompt temperature parameter\footnote{The prompt temperature parameter controls the ``creativity" or randomness of the text generated by the LLM. It ranges from 0 to 2, where higher values (e.g., 1) produce more diverse and creative responses, while lower values (e.g., 0.1) yield more deterministic and focused outputs.} (A1, A5, A11). Study A11 further experimented with different LLMs to build confidence in the responses and selected the model output that scored highest on a predefined criteria.

In terms of the prompt components used, in three studies (A4, A6, A8), examples were provided in the prompt to guide the model in conducting a task or sub-task. In A6 and A8 the authors provided the model with an example of initial code generation, guiding it to observe and replicate the implied logic. On the other hand, in A4, multiple examples were provided by the researcher to the model to generate initial codes following the few-shot strategy. These examples were taken from the test data, annotated by the researcher and inputted to the model.

Seven studies (A1, A3, A5-A6, A8, A10-A12) framed the task for the model in particular ways. Five (A1, A5-A6, A8, A11) explicitly defined the role they wanted the model to assume, often tailored to the ITA context, to help refine the model’s responses by aligning with qualitative research methodologies. These would include explicitly requiring the model to play the role of a thematic/qualitative research assistant or expert text or qualitative data analyst/bot. Seven studies (A1, A3, A6, A8, A10-12) included additional details on the research setting, highlighting the investigated research problem and the analytical method being used for the analysis (i.e., ITA). Furthermore, three studies (A6, A8, A10) specified the type of input data (e.g., interview transcripts)  and one study (A12) additionally indicated the input file format (JSON) in the inputted prompt.

Ten (A1-A2, A4-A9, A11-A12) out of the 12 studies provided structured guidance on how the model should present the ITA outputs. Researchers frequently specified formats for initial codes and themes, including elements such as titles, descriptions, and example quotations from the input data to support each code or theme. In some studies, researchers explicitly defined the number of codes or themes they wanted the model to generate (A1, A7, A9, A11). However, most studies avoided restrictions on the number of outputs to align with traditional ITA approaches where human analysts focus on the identification of high-quality codes and themes rather than their quantity. Furthermore, some studies (A4, A6-A8, A12) required the model to generate ITA outputs in a specific JSON file format, while the rest did not specify an output file format. In A11, the authors also required the model to provide step-by-step reasoning chain  as it generates themes in the analysis.

In addition, brief commentary is provided in five studies (A4, A6, A10-A11) on the issues the researchers faced in the specification of the language used to define prompts. The main concerns emanate from the length of context window, which does not allow all the interview transcripts to be fed to the model at once. Thus, in these studies, an interview transcript is segmented into equal-length word chunks to make it accessible for the model.  A6 and A12 also point towards the importance of using a balanced prompt language that aligns with both the model's vocabulary and the terminologies used in ITA research.

\subsubsection{DEQ6: Does the study explain how model responses were refined?}
The response refinement process is briefly discussed in studies A4–A5, A7, and A11–A12. In A4, a conversational approach was proposed and implemented, where researchers iteratively evaluated the model’s responses, suggested modifications with reasoning, and engaged in a back-and-forth process until mutual agreement between the researcher and the model was reached. A5 refined the initial prompt by including contrastive examples, i.e., a set of examples to describe the incorrect reasoning chain incorporated in these examples is provided to the model to guide it toward the final ITA outcome. In A7 and A12, prompts were modified by adjusting the number of expected themes and rephrasing the prompt language, respectively. Similarly, A11 refined the prompt by rephrasing the response instructions to improve model outputs.

\subsection{Analysis}

Although the application of LLMs to support ITA is still in its early stages, it is gaining traction across various academic disciplines including healthcare, education, and transportation.  These models are being applied to thematically analyze diverse qualitative data, such as interviews, open-ended survey responses, and social media posts. Multiple studies demonstrated that LLM-assisted thematic analysis yields results comparable to traditional methods, thereby boosting confidence in its application. However, the authors of the selected studies suggest that the integration of LLMs into ITA should only be advanced as a supporting tool, for example to function as an additional investigator in investigator triangulation, and strongly caution against the replacement of human researchers by LLMs. According to these studies, expert intervention remains essential to effectively prompt the model, select appropriate codes and themes, and ensure that model responses align with the research objectives. 

ITA researchers have adopted various approaches to PE in their analyses. While some completed the entire ITA process in a single step, others employed multiple prompts to achieve their objectives. Most studies focused on output formatting and prompt framing, with only a few incorporating examples to guide the model's responses. When refining model outputs, researchers have primarily relied on human interpretation of the generated results to provide feedback and enhance the model’s responses. The discussion on how ITA researchers have approached PE to support their analysis leads to the identification of four key steps in the PE process. These include (i) selecting a task completion strategy, (ii) developing prompt templates, (iii) crafting and executing prompts, and (iv) iteratively refining model responses. These four steps are examined further in the following section.

\begin{table*}[]
\resizebox{\textwidth}{!}{
\begin{tabular}{@{}c|p{1.75cm}p{1.75cm}|p{1.75cm}p{1.75cm}p{1.75cm}|p{1.75cm}p{1.75cm}@{}}
\toprule
\multirow{2}{*}{\textbf{ID}} & \multicolumn{2}{c|}{\textbf{Task Completion Strategy}} & \multicolumn{3}{c|}{\textbf{Prompt Format}}       & \multicolumn{2}{c}{\textbf{Response Refinement}} \\ \cmidrule(l){2-8} 
                    & \multicolumn{1}{c}{Single-prompt}          & \multicolumn{1}{c|}{Multi-prompt}          & \multicolumn{1}{c}{Exemplars} & \multicolumn{1}{c}{Framing} & \multicolumn{1}{c|}{Response instructions} & \multicolumn{1}{c}{Human-assisted}     & \multicolumn{1}{c}{Self-refinement}    \\ \midrule
A1  & \multicolumn{1}{c}{$ \times $} &   &   & \multicolumn{1}{c}{$ \times $} & \multicolumn{1}{c|}{$ \times $} &   &  \\
A2  & \multicolumn{1}{c}{$ \times $} &   &   &   & \multicolumn{1}{c|}{$ \times $} &   &  \\
A3  & \multicolumn{1}{c}{$ \times $} &   &   & \multicolumn{1}{c}{$ \times $} &   &   &  \\
A4  &   & \multicolumn{1}{c|}{$ \times $} & \multicolumn{1}{c}{$ \times $} &   & \multicolumn{1}{c|}{$ \times $} & \multicolumn{1}{c}{$ \times $} &  \\
A5  &   & \multicolumn{1}{c|}{$ \times $} &   & \multicolumn{1}{c}{$ \times $} & \multicolumn{1}{c|}{$ \times $} & \multicolumn{1}{c}{$ \times $} &  \\
A6  &   & \multicolumn{1}{c|}{$ \times $} & \multicolumn{1}{c}{$ \times $} & \multicolumn{1}{c}{$ \times $} & \multicolumn{1}{c|}{$ \times $} &   &  \\
A7  &   & \multicolumn{1}{c|}{$ \times $} &   &   & \multicolumn{1}{c|}{$ \times $} & \multicolumn{1}{c}{$ \times $} &  \\
A8  &   & \multicolumn{1}{c|}{$ \times $} & \multicolumn{1}{c}{$ \times $} & \multicolumn{1}{c}{$ \times $} & \multicolumn{1}{c|}{$ \times $} &   &  \\
A9  & \multicolumn{1}{c}{$ \times $} &   &   &   & \multicolumn{1}{c|}{$ \times $} &   &  \\
A10 &   & \multicolumn{1}{c|}{$ \times $} &   & \multicolumn{1}{c}{$ \times $} &   &   &  \\
A11 &   & \multicolumn{1}{c|}{$ \times $} &   & \multicolumn{1}{c}{$ \times $} & \multicolumn{1}{c|}{$ \times $} & \multicolumn{1}{c}{$ \times $} &  \\
A12 & \multicolumn{1}{c}{$ \times $} &   &   & \multicolumn{1}{c}{$ \times $} & \multicolumn{1}{c|}{$ \times $} & \multicolumn{1}{c}{$ \times $} &  \\ \bottomrule 

\end{tabular}}
\vspace{0.2cm}
\caption{Literature Review Findings.}\label{tab:sims500res}
\end{table*}

\section{The Four-Step PE Process}

\label{Sec 4}

\begin{figure}
    \centering
    \footnotesize
    \begin{tikzpicture}[node distance=1.8cm]

    \node (step1) [process] {Step 1: Devise task completion strategy};
    \node (step2) [process, below of=step1] {Step 2: Create prompt template};
    \node (step3) [process, below of=step2] {Step 3: Define and execute prompts};
    \node (decision1) [decision, below of=step3, aspect = 1.5, yshift=-0.5cm] {Are model responses satisfactory?};
    \node (step4) [process, left of=decision1, xshift=-3cm] {Step 4: Refine responses};
    \node (step5) [process, below of=decision1, yshift=-0.5cm] {Final model responses};

    \draw [arrow] (step1) -- (step2);
    \draw [arrow] (step2) -- (step3);
    \draw [arrow] (step3) -- (decision1);
    \draw [arrow] (decision1.west) -- (step4.east) node[midway, above] {No};
    \draw [arrow] (step4.north) |- (step1);
    \draw [arrow] (step4.north) |- (step2);
    \draw [arrow] (step4.north) |- (step3);
    \draw [arrow] (decision1.south) -- (step5) node[midway, right] {Yes};
    

    \end{tikzpicture}
    \caption{Prompt Engineering (PE) Process.}
    \label{fig:PE_phases}
\end{figure}
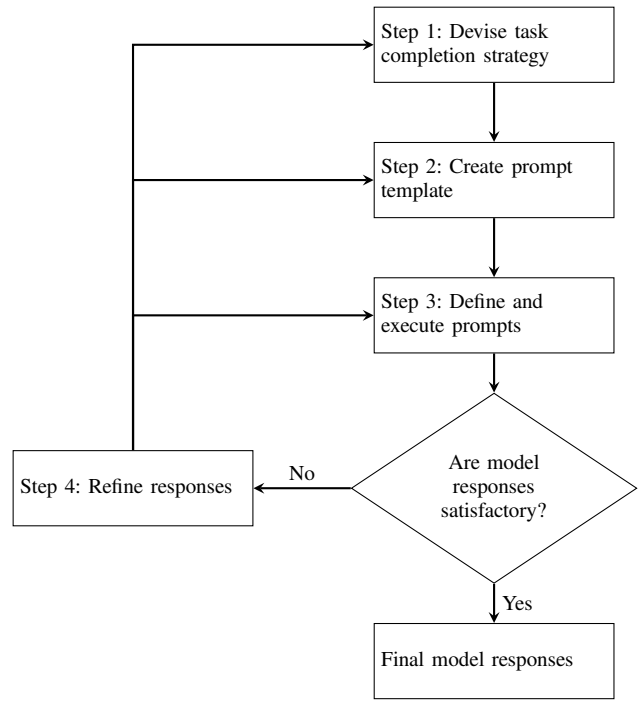

Built on the understanding of how ITA researchers have approached PE in the reviewed studies, four key steps in the PE process can be identified (see Figure \ref{fig:PE_phases}). Accordingly, ITA researchers practicing PE first devise a task completion strategy, i.e., a high-level plan for how they want to use LLMs to support their analysis. Next, they create prompt templates guided by the chosen task completion strategy. Task-specific prompts are then defined and executed using the selected LLM. Finally, the model’s responses are evaluated and, if necessary, any of the previous steps are iteratively revised until the desired outcomes are achieved.

\subsection{Step 1: Devise task completion strategy}
The first step in the PE process is to strategically devise a plan to conduct LLM-assisted ITA. In this step the researcher provides a high-level blueprint that outlines how they intend  to leverage LLMs to effectively complete their analysis. This involves answering questions such as whether to use LLMs to execute the analysis at once or to employ more methodological approaches to conduct the analysis by prompting the model multiple times. In this context, the two important task completion strategies highlighted in Section \ref{Sec 3} include (i) single-prompt strategy and (ii) the multi-prompt strategy. 

\subsubsection{Single-prompt strategy}
The single-prompt strategy refers to the simple application of a single well-structured prompt to effectively guide the LLM to complete ITA. Here, the researcher provides the model with few-shot examples, task framing, and response instructions, all in one prompt. 

\subsubsection{Multi-prompt strategy}
The multi-prompt strategy refers to the application of multiple prompts to complete ITA. When applying the multi-prompt strategy, the ITA researcher outlines the conditional logic, the prompt execution frequency, or other architectural considerations essential for effective prompting. Two important multi-prompt strategies discussed in Section \ref{Sec 3} include the decomposition strategy and the ensembling strategy.

\begin{itemize}
    \item \textbf{Decomposition strategy}
\end{itemize}

This approach involves the application of multiple prompts to complete ITA where the researcher explicitly breaks down the ITA into multiple sub-tasks (e.g., by explicitly decomposing ITA into six sub-tasks following Braun and Clarke's method). Depending on the researcher’s specific requirements, these sub-tasks can be further broken down into sub-subtasks, with the decomposition process continuing as needed and per the researcher's discretion. 
    
\begin{itemize}
    \item \textbf{Ensembling strategy} 
\end{itemize}

The ensembling approach involves application of multiple prompts to generate model responses, then aggregating these responses to select the best outcome. In this process, each response is evaluated on a pre-determined criterion where the evaluation may be conducted by human annotators or the researcher may request the LLM to self-evaluate. 


\subsection{Step 2: Create Prompt Templates}
In this step, ITA researchers create a template for each prompt that is required by the task completion strategy. Here the researcher specifies what prompt elements they intend to include in their inputs to the model. As discussed in Section \ref{Sec 3}, the prompt templates used by ITA researchers frequently comprises a combination of the following three formats.

\begin{itemize}
    \item \textbf{Exemplar prompt format}
\end{itemize}
With the exemplar prompt format (also known as few-shot prompt format) ITA researchers provide the model with a set of directive-response examples to induce an understanding of how to conduct ITAs. The goal is to enable the model to infer the underlying logic required to conduct the analysis without the need to retrain the model or update its weights. Here, the ITA researcher can provide, for instance, examples of initial code generation or theme identification to the model. This approach allows ITA researchers to provide the model with examples that may contradict the model's training data to focus on a particular outlier phenomenon, and to guide the model in less-known directions. 

\begin{itemize}
    \item \textbf{Framing prompt format}
\end{itemize}
Prompt framing format enable a more nuanced understanding of the analysis in LLM-assisted ITA. Using this format, the researcher explicitly requests that the model take particular roles while conducting the analysis, or provides emotional instructions to emphasize phrases that are of psychological importance to humans. Prompt framing also includes provision of additional background information to the model, for example, the description of the research question or the type of study or input data (interview, survey, etc.) that the model is being directed to analyze.  

\begin{itemize}
    \item \textbf{Response instruction prompt format}
\end{itemize}
ITA researchers can also include instructions in the prompt template that direct the model to generate its responses in a specific form. This involves specification of the desired response style (e.g., tone, genre), the format of the output file (e.g., MS Word, Excel, PDF) or the format of the model responses (e.g., themes presented in a table format). In addition, some response formats require the model to produce its step-by-step chain-of-thought (CoT), to explain how it arrived at its response. This is done by appending a thought inducer such as ``Let's think step-by-step" to the prompt that asks the model to provide the logic it implied in its response.

\subsection{Step 3: Define and Execute Prompts}
The creation of prompt templates is followed by the definition and execution of task-specific prompts. This step involves crafting clear, concise instructions that are tailored to ITA context. Here, the ITA researcher ensures that the language used in the prompt aligns with the analysis’ objectives and is also compatible with the model's vocabulary. If there are conflicting terms or terminology that the model may not fully understand or which the LLM may misunderstand, the researcher makes adjustments to match the model’s language. 

Another important factor is the effective use of context window, i.e., the number of tokens that a model can process at once, where a token is the smallest text unit processed by an LLM, often smaller than a word, such as sub words or characters. Every LLM has token limits (for example, 2048 or 1 million tokens) for prompts and model responses, and exceeding these can lead to incomplete processing. Thus, when defining prompts, ITA researchers cater to the limits imposed by the context window of the model used. After the prompt is defined, it is executed through input into the model.

\subsection{Step 4: Refine Model Responses}

The next critical step in the PE process is to iteratively refine the LLM’s responses. Here, the ITA researcher has a choice to either modify the task completion strategy, alter the prompt template, or simply change the language used in the original prompt. Often, multiple such iterations are required to achieve the researcher's desired responses. Each iteration requires evaluation of the model’s initial outcomes using a scoring mechanism and providing effective feedback. Determined by the feedback provision method, the two important response refinement approaches include:

\begin{itemize}
    \item \textbf{Human-assisted refinement} 
\end{itemize}
The more commonly used response refinement approach implemented by ITA researchers involves human-in-the-loop to provide feedback on the model-generated responses. To achieve this, the researcher relies on their own judgment or consults subject-matter experts. The conversational capabilities of LLMs also enable researchers to refine responses by interacting with the model itself. In this setting the ITA researcher incorporates the outcomes of the initial prompt execution into subsequent prompts to allow for an iterative and collaborative improvement of response quality.

\begin{itemize}
    \item \textbf{Self-refinement} 
\end{itemize}
Although ITA researchers have largely avoided using LLMs for self-refinement, this presents an additional strategy that can be leveraged. In this approach, the model is prompted to evaluate and improve its own responses based on a structured assessment. This involves instructing the model to assess the reasonableness of its outputs, identify potential shortcomings, suggest logical modifications, and refine its responses accordingly.

\section{A Discussion on the State-of-the-Art PE Techniques}

\label{Sec 5}


Having outlined the four key steps in the PE process, this section references the findings from Section \ref{Sec 3} and discusses the application of more recent and advanced PE techniques and guidelines to support the execution of these steps in the ITA context. The aim is to highlight the techniques that ITA researchers have already employed and explore additional methods that can further enhance their LLM-assisted analyses.

\subsection{Task Completion Strategy}

\subsubsection{Single-prompt strategy}
Multiple ITA researchers employed the single-prompt task completion strategy to conduct the complete ITA with one prompt. Here, the prompts used by the ITA researchers varied according to the different prompt components that they incorporate and the task-specific definition of the prompt. Although no PE technique specific to only the single-prompt task-completion strategy can be highlighted, multiple PE techniques mentioned in the following sub-sections can be used by ITA researchers in this setting.

\subsubsection{Multi-prompt strategy}

\begin{itemize}
    \item \textbf{Decomposition Strategy}
\end{itemize}
Many ITA researchers broke down the ITA process into multiple sub-tasks, following Braun and Clarke's ITA process, and conducted only some steps utilizing LLMs. This approach closely aligns, though not entirely, with the decomposition PE technique \cite{khot2022decomposed}, in which researchers sequentially divide a complex task into logical sub-tasks to guide the task completion step-by-step. The decomposition can be hierarchical or recursive, and can make use of an external LLM to complete sub-tasks in parallel. For example, the application of this technique to perform ITA can be achieved using the following sequence of prompts:

\begin{quote}
    \textit{Prompt 1}: Read the following interview responses and summarize the key ideas. Then, extract key phrases as initial codes.\\
    \textit{Prompt 2}: Now, group similar codes into potential themes, ensuring each theme has a clear name and definition.\\
    \textit{Prompt 3}: Finally, check for overlapping or contradictory themes and refine them before presenting your final thematic analysis.
\end{quote}

Another approach that can add incremental depth to ITA is the least-to-most PE technique \cite{zhou2022least}. In this method, tasks are structured according to their increasing level of complexity to ensure that the model first solves fundamental sub-tasks before it tackles the more challenging ones. This technique can improve rigor in ITA by gradually delving into the depths of the analysis. For example, ITA researchers can use the following sequence of prompts to uncover the hidden patterns in the text: 

\begin{quote}
    \textit{Prompt 1}: Given the interview data below, start by identifying the most explicitly mentioned themes.\\
    \textit{Prompt 2}:Now, look for secondary themes that may not be immediately obvious but are still clear.\\
    \textit{Prompt 3}:Finally, explain how these themes relate to each other and their broader significance.
\end{quote}

\begin{itemize}
    \item \textbf{Ensembling strategy}
\end{itemize}

When experimenting with prompts, ITA researchers typically tuned the prompt temperature setting to generate multiple candidate responses, which were then aggregated through visual inspection. In this process, the themes generated from each execution of the prompt with non-zero temperature setting were evaluated by human annotators and the themes that were more reasonable were selected. 
    
This approach is similar to the universal self-consistency PE technique \cite{chen2023universal}, where the model is prompted multiple times with the same input with a non-zero prompt temperature setting. However, in the self-consistency PE technique, all the model-generated responses are concatenated and are inputted back to the model, which selects the most consistent responses. Thus, the themes generated by the prompt executions can be included in a subsequent prompt for the model to select the most appropriate responses. 

In addition, ITA researchers can also apply the demonstration ensembling PE technique \cite{khalifa2023exploring}. According to this technique, the researcher creates multiple few-shot prompts, each including a distinct set of examples. These prompts are then executed and the model responses are aggregated to generate a final response. In the case of thematic analysis, for instance, from the master set of theme identification examples, multiple subsets of distinct examples will be included in the various prompts executed by the researcher. The model's responses from these executions will then be aggregated to determine the final response, i.e., the final set of themes in the analyzed text corpus. 

Another approach that ITA researchers can adopt is multi-persona prompting \cite{gao2023self}. This method leverages the LLM’s ability to maintain consistent character traits and enables the model to ``wear different hats" during interactions. For instance, researchers can prompt the model to perform ITA from the perspective of both a novice and an expert qualitative data analyst, thereby adding depth to analysis by incorporating diverse viewpoints and perspectives.

\subsection{Prompt template}
\begin{itemize}
    \item \textbf{Exemplar prompts} 
\end{itemize}

ITA researchers have generally avoided the use of examples in their prompts although increasing the number of examples has shown improvements in LLM-generated responses across various tasks \cite{brown2020language}. Except in \cite{dai2023llm}, the remaining studies relied on the provision of a single such example. Interestingly, the examples included in the prompts to the model were all furnished by the researcher.

Yet another approach that ITA researchers can incorporate is to request the model to generate its own examples before inputting all or a subset of these examples back to the model to induce the logical completion of ITA. This PE technique, called the self-generated, in-context learning \cite{kim2022self}, reduces the reliance on external demonstrations. Consider the case when a researcher intends to perform ITA on an interview transcript with the support of LLMs. In this case, ITA researchers can request the model to self-generate theme generation examples by asking it to: ``please come up with three new and creative examples of theme generation in inductive thematic analysis." These examples will then be referenced by the model when carrying out an ITA. 

The model can also be instructed to generate few-shot reasoning chain examples to complete a task using the analogical PE technique \cite{yasunaga2023large}. This technique draws inspiration from a psychological notion (analogical reasoning) where people use pertinent prior experiences to solve new problems. An example of ITA researchers implementing the analogical PE technique is provided below. Note how the prompt requires the model to build on its prior experience to complete the task:
    
    \begin{quote}
    Your task is to perform thematic analysis. When presented with the interview data, recall relevant and similar data examples, and proceed to generate themes in those examples and provide a step-by-step reasoning. Afterwards, identify themes in the interview data given below.
    \end{quote}
    

\begin{itemize}
    \item \textbf{Framing prompts}
\end{itemize}

According to Section \ref{Sec 3}, ITA researchers often utilized the framing prompt format. In this setting, some studies specified the role the model was required to assume while others focused on providing the task context to the model. 
    
An additional PE technique, called the emotional PE technique, relates to the inclusion of phrases with psychological relevance to humans within the prompt. \cite{li2023large} demonstrated that explicitly including emotional stimuli in the prompt improved LLM performance on open-ended text generation tasks. This can be achieved by ITA researchers, for example, by appending the prompts with statements such as ``This is the most important task to my thesis data analysis" or ``make sure you are confident in your repsonses" to the ITA prompt. This strategy encourages the LLM to recognize and respond to the emotional cues that provide humans with a unique advantage in conducting ITA.

In other cases, ITA researchers may also consider instructing the model to adopt a particular mood or style. This technique, also known as style prompting, can be used to control various stylistic elements of the generated text, such as writing style, tone, mood, plot, genre, and pace \cite{lu2023bounding}. In the ITA context, researchers can pursue the application of style prompting by including statements such as ``avoid overly simplistic theme description, and instead, aim for a more nuanced explanation of the themes", and ``for each theme, please identify any emotional tones that accompany it" in their prompts.

\begin{itemize}
    \item \textbf{Response instruction prompts}
\end{itemize}
    
ITA researchers have comprehensively included details on how the model should generate its responses. The concentration has been on the output structure and format, while some researchers have also required the model to append its reasoning process to the generated responses. In this domain, ITA researchers can also experiment with other variations of thought inducers such as ``first let’s think about it logically” or ``let’s work this out in a step-by-step way to be sure we have the right answer” \cite{schulhoff2024prompt} to evaluate the model responses.

In \cite{mathis2024inductive}, the authors provided examples of incorrect logic that the model should not follow when responding to the task. In the literature, this technique is referred to as the contrastive CoT PE technique \cite{chia2023contrastive}. Here, both positive and negative task-specific demonstrations are provided to the LLM to enhance its reasoning capabilities. ITA researchers can accomplish this, for example, by asking the model to first generate themes with their reasoning chain in a corpus multiple times by varying the prompt temperature setting. The model-generated themes and their reasoning chain may then be evaluated by the researcher, and the annotated examples with corrections to the incorrect logic implied in the initial responses can serve as few-shot examples for the model to complete ITA.
    
A more advanced PE technique that builds on incorporating the model’s reasoning chain in its responses is the active PE technique proposed in \cite{diao2023active}. Active prompting enhances LLM reasoning through self-generated few-shot reasoning chains, uncertainty estimation, and human annotation. For instance, an ITA researcher aiming to conduct initial coding on ten open-ended survey questions using this approach would first provide a few (e.g., three) examples of initial coding with their reasoning chains to the model. The model would then be queried on new cases from the remaining survey questions. Responses with high uncertainty would be identified and annotated by the researcher. These annotated responses, along with their corrected reasoning chains, would then be incorporated as few-shot examples to guide the coding of the remaining survey questions, instructing the model to generate a similar reasoning chain in its responses.

\subsection{Prompt Definition and Execution}
Prompt definition strictly depends on the specific task that the researcher wants the model to accomplish. Due to this subjectivity, no specific PE technique can be identified to support this PE phase. However, general guidance on prompt definition can be found in multiple studies \cite{marvin2023prompt, dair2025, openAi2025, giray2023prompt}, and is synthesized below:   
\begin{itemize}
    \item \textbf{Unambiguity:} The prompt language should be clear, concise, and strictly relevant to both the task and the model's capabilities. It should provide sufficient context to eliminate potential confusion and ensure that the model accurately interprets and executes the request. 
    \item \textbf{Precision:} A balance between specificity and generality should be sought. The goal should be to ensure that the model remains flexible enough to generate diverse and contextually relevant responses while still adhering to the intended guidelines. 
    \item \textbf{Constructiveness:} Rather than instructing the model on what to avoid, it is more effective to guide it toward the desired actions and behaviors. 
    \item \textbf{Ethical Considerations:} Prompts should adhere to ethical standards by ensuring fairness, neutrality, and inclusivity in language and intent. It is essential to eliminate biased phrasing and avoid any pre-conceived notions or assumptions that could influence the model’s responses.
    \item \textbf{Delimiters:} The effective use of delimiters helps clearly distinguish different sections within a prompt, enhancing readability and reducing ambiguity. Common delimiters such as $``````$ (triple back ticks) and \#\# (double hash marks) are frequently used to separate instructions, examples, or specific input data.
    \item \textbf{Maximum Length:} Defining the desired length of model responses helps ensure conciseness, relevance, and adherence to task requirements. By specifying a word count, character limit, or sentence constraint, the prompt can guide the model to generate responses that are appropriately detailed without being overly verbose or insufficiently informative.
\end{itemize}

\subsection{Response Refinement}

\begin{itemize}
    \item \textbf{Human-assisted refinement} 
\end{itemize}
As outlined in Section \ref{Sec 3}, ITA researchers often employ the prompt paraphrasing PE technique \cite{jiang2020can} to enhance model-generated responses by partially rewording or rephrasing them. These researchers rely on their own judgment to continue the paraphrasing process until the desired model response is achieved. Paraphrasing can also include rephrasing the thought inducer used in the prompt. Instead of ``Let's think step-by-step," the thread of thought PE technique \cite{zhou2022large} improves the model response by using thought inducers such as ``Walk me through this context in manageable parts step by step, summarizing and analyzing as we go." This technique is shown to be more effective in handling long chaotic text such as interview transcripts. Rephrasing can also be achieved by ITA researchers by implementing the system-2-attention PE technique \cite{weston2023system}. According to this technique, the model is required to remove any unrelated information and rewrite the prompt before passing the revised prompt into the LLM to retrieve the improved responses. This can be achieved by simply appending the prompt with statements such as the following: ``Read the prompt and rewrite it removing any unrelated information."

\begin{itemize}
    \item \textbf{Self-refinement}   
\end{itemize}
Section \ref{Sec 3} highlights that ITA researchers have yet to utilize LLMs for self-evaluation and refinement of their responses. Two relevant PE techniques that can aid in this process are the self-calibration PE technique \cite{kadavath2022language} and the self-refine PE technique \cite{madaan2024self}.

With the self-calibration PE technique, an ITA researcher analyzing an interview transcript, for instance, can first prompt the model to identify themes within the text. The model's response will then be fed back into the model, prompting it to assess its confidence in the generated themes and the reasoning behind them. The model's responses to the second prompt enable the researcher to modify the prompt accordingly. The self-refine PE technique, on the other hand, will require the model to improve its initial responses in addition to just providing feedback on its responses. This iterative refinement continues for a predetermined number of cycles.

\begin{table*}[h]
\resizebox{\textwidth}{!}{%
\begin{tabular}{@{}p{2.25cm}p{2.25cm}|p{2.25cm}p{2.cm}p{2.5cm}|p{2.25cm}p{2.25cm}@{}}
\toprule
\multicolumn{2}{c|}{\textbf{Task Completion Strategy}} &
  \multicolumn{3}{c|}{\textbf{Prompt Format}} &
  \multicolumn{2}{c}{\textbf{Response Refinement}} \\ \midrule
Decomposition & Ensembling               & Exemplar   & Framing & Response Instructions & Human-assisted     & Self        \\ \midrule
Decompose \cite{khot2022decomposed} &
  Universal self consistency \cite{chen2023universal}&
  Self-generated in-context learning \cite{kim2022self}&
  Emotional \cite{li2023large}&
  Contrastive CoT \cite{chia2023contrastive}&
  Prompt paraphrasing \cite{jiang2020can}&
  Self calibration \cite{kadavath2022language}\\
Least-to-most \cite{zhou2022least}& Demonstration ensembling \cite{khalifa2023exploring}& Analogical \cite{yasunaga2023large}& Style \cite{lu2023bounding}  & Active  \cite{diao2023active}              & System-2-attention \cite{weston2023system} & Self-refine \cite{madaan2024self}\\
              & Multi-persona \cite{gao2023self}            &            &         &                       &                    &             \\ \bottomrule
              \vspace{0.2cm}
\end{tabular}%
}
\caption{State-of-the-art PE Techniques.}
\label{tab:my-table}
\end{table*}

\section{Summary of Contributions and Future Directions}

\label{Sec 6}

This paper presents a comprehensive analysis of the use of LLMs to support ITA. It provides a structured approach to investigate how ITA researchers have incorporated PE in their analyses and discusses the application and integration of advanced PE techniques into ITA practices. 

\subsection{Contributions}
The key contributions of this article include:

\begin{itemize}
    \item \textbf{Mapping of the field} 
\end{itemize}
Through a systematic review of the literature, the study provides an overview of the use of LLMs for ITA and the particular PE techniques used by ITA researchers. This review explored key dimensions including, (i) the study field in which LLMs have been applied to support ITA, (ii) the primary objective of these studies and the identified limitations, (iii) the LLM used and the data sources employed, (iv) metrics used to evaluate the effectiveness of LLM application, (v) how PE is approached, (vi) the key prompt elements used, and (vii) the response refinement strategy employed. The results highlight the recent nature of explorations in this field and the opportunistic approach pursued by ITA researchers to define prompts and refine model responses.

\begin{itemize}
    \item \textbf{Structuring of the PE process} 
\end{itemize}    
The potential of LLMs to reduce the time- and cost-related challenges of ITA has been widely explored by researchers. However, the use of LLMs in ITA has often been opportunistic, with prompts being applied in an ad hoc manner. This behavior compromises replicability, objectivity, and transparency of the analysis and highlights an urgent need for a structured approach to PE in ITA. Built on the insights generated from the literature review, this article highlights four key steps in the PE process that have been used by ITA researchers to organize their thought process and enhance the effectiveness of their prompts. The four steps include: developing a task completion strategy, designing prompt templates, defining and executing prompts, and refining the model responses until the desiderata are met.

\begin{itemize}
    \item \textbf{Application of Advanced PE Techniques to ITA research} 
\end{itemize}
The discussion in the paper also highlights advanced PE techniques and how ITA researchers can incorporate these techniques into their analyses. It is contended that an examination of the application of these advanced techniques can further strengthen the ITA community's confidence in the use of LLMs by improving the quality of the model generated responses.

\subsection{Future research directions}
The findings of this study highlight various research opportunities presented below:

\begin{itemize}
    \item \textbf{Strengthening ethical research} 
\end{itemize}
While the use of LLMs and the various PE techniques have been extensively explored in this article, it does not imply that the application of LLMs to assist ITA is flawless. LLM-assisted ITA is still in its early stages and deliberate efforts to study the ethical considerations regarding the use of these models are imperative. Several questions can be explored, such as, ``Does the use of LLMs introduce, amplify, or mitigate the subjective biases that accompany ITA? What are the ethical concerns regarding the use of sensitive human-subject data? How can researchers prevent overreliance on LLMs? What ethical safeguards should be placed to enable the ethical use of LLMs to support data analysis?" Exploring these questions is both essential and intriguing. 

\begin{itemize}
    \item \textbf{Establishing clarity on the PE process}
\end{itemize}
The various studies analyzed in this work report a high degree of similarity between LLM-assisted and traditional ITA. However, the limited reporting on the PE methods used in these studies challenges the validity of these claims. Since the same analysts who conducted traditional ITA also formulated the prompts for LLM-assisted ITA, there is a strong possibility of subjective bias in PE. Thus, there is significant need for ITA researchers to establish agreed-upon procedures that can be adhered to when using LLMs. This is important because establishing such procedures will provide clarity and will enable reliability in the results, factors that are key to establishing quality in qualitative research. Although this work provides a structured four-step PE process, in-depth investigations related to each of the four phases will be both necessary and timely. In addition, the application of the process and the consequent time- and cost-saving must be investigated to better appreciate the benefits of integrating LLMs in ITA research using this process.

\begin{itemize}
    \item \textbf{Evaluating the use advance PE techniques}
\end{itemize}
While multiple advanced PE techniques are highlighted in this work and their application in the ITA context is briefly discussed, a systematic investigation of their use in ITA research still remains unexplored. The implementation of the various techniques in different research context can help in understanding their contextual suitability.




\end{document}